\documentclass[aps,prl,twocolumn,showpacs,superscriptaddress,groupedaddress]{revtex4}  
\usepackage{graphicx}  
\usepackage{dcolumn}   
\usepackage{bm}        
\usepackage{amssymb}   
\usepackage{amsmath}
\usepackage{CJK}  

\begin{document}
\begin{CJK*}{UTF8}{}


\title{Generating Steady Topologically Non-Trivial Artificial Spin Texture with Cold Atoms}
\author{
Junyi Zhang
}
\email[]{junyi.zhang@ens.fr}
\affiliation{D\'epartement de Physique, l'\'Ecole Normale sup\'erieure, 24 rue Lhomond, 75005, Paris, France}

\date{\today}

\begin{abstract}
  In this article we proposed a scheme to generating steady topologically non-trivial artificial spin texture in cold atom systems.  An example of generating a texture of charge one skyrmion with Laguerre-Gaussian beam was given.  It provides a scheme for studying skyrmion excitations of quantum Hall ferromagnetism in cold atom systems.
\end{abstract}


\maketitle
\end{CJK*}

Cold atoms provide an on-table platform for studying a wide range of physical phenomena.  It is not only important for atomic-optical physics, but also used as a powerful quantum simulator to study condensed matter physics, quantum information and even high-energy physics~\cite{Bloch2012}.  Optical lattice were used for realizing important condensed matter models~\cite{Bloch2008}.  One may also generate artificial gauge fields for neutral atoms~\cite{Dalibard2011}.  Optical flux lattice provides a possible scheme to reach a regime of high flux density~\cite{Cooper2011} to study quantum Hall states.  Some proposals have been realized experimentally (for example~\cite{Aidelsburger2013}\cite{Atala2013}).

In real quantum Hall state, electrons are polarized to form ferromagnetic states due to the exchange interaction of Coulomb force.   Single electron excitation from $\nu=1$ quantum Hall states also twists the spin texture of electrons and form skyrmion~\cite{Sondhi1993}.  Skyrmion was first introduced in 1961 as a topological soliton to describe nuclei~\cite{Skyrme2013}.  It has been found to have close connections to superconductivity, quantum gravity etc.  Skyrmion in spinor Bose-Einstein condensates ave been studies theoretically~\cite{Zhai2003}, and realized in experiments~\cite{Leslie2009}.  However it is found to be energetically unstable.

In this article, we describe a scheme to generating steady topologically non-trivial artificial spin texture with light induced effective magnetic field.  This effective magnetic field is different from most proposals of artificial gauge fields that emerges as Berry curvature.  It couples directly to the internal states as an effective Zeeman field orienting the ``spin'' to form the texture.  Combining with the techniques of generating artificial magnetic field with optical flux lattice, we may study skyrmion excitation of quantum Hall ferromagnetism cold atom system.  We also gave an example generating a skyrmion of charge $c=\pm1$ in cold atom system with Laguerre-Gaussian beam.  This realization of spin textures in BEC does not depend on the stability of real skyrmion in spinor condensates.  Therefore it provides a better simulation for studying spin textures in quantum Hall states with cold atoms.

Let us first consider an electron confined in $x$-$y$ plane moving adiabatically in an external magnetic field.  Suppose that at origin, the magnetic field $\mathbf{H}$ is in $+z$ direction, and the electron is in the ground state spin down $|\downarrow\rangle$.  Along any radial direction, the magnetic field tilts towards the radial direction, lies in  $x$-$y$ plane at circumference of unit circle, flips downwards outside the unit circle, and points $-z$ direction at infinity.
Since the magnetic field does not vanish at any point, the adiabatic theorem guarantees that the electron should always stay in the ground state, i.e. its spin is aligned counter to the direction of magnetic field.  This is an simple realization of skyrmion of charge $c=\pm1$.  Mathematically, the configuration of the spin is homotopical to stereographic projection $f: S^2 \rightarrow S^2$ by one-point compactification of $x$-$y$ plane to $S^2$ (identifying $\infty$ to a pole).

To realize such a spin texture with cold atoms, we need a ``magnetic field'' to align the ``spin'' of the atom.  Consider a Two-Level Atom (TLA) with internal states $|e\rangle$ and $|g\rangle$.  Unitary transformation group on the internal Hilbert space is $SU(2)$ that enable us to simulate the spin with TLA.  The Hamiltonian of the internal degree of freedom of TLA is
\begin{equation}\label{eq:HamAtomInternal}
  H_{\text{atom}}=\frac{1}{2}\hbar\omega_{eg}\sigma_z
  =\frac{\hbar}{2}\begin{pmatrix} \omega_{eg}  &  0\\
  0 & -\omega_{eg} \end{pmatrix},
\end{equation}
where $\omega_{eg}$ is the energy difference of two internal states.
We considering the dipole coupling of the atom to external field, the interacting Hamiltonian is
\begin{equation}\label{eq:HamDipoleCoupling}
  H_{\text{int}}=-\hat{\mathbf{D}}\cdot\mathbf{E}
  =-\hbar
  \begin{pmatrix} 0  & \mathbf{d}\\
  \mathbf{d} & 0 \end{pmatrix} \cdot \mathbf{E},
\end{equation}
where $\hat{\mathbf{D}}=\begin{pmatrix} 0  & \mathbf{d}\\
\mathbf{d} & 0 \end{pmatrix}$ is the dipole operator acting on the internal space of the atom and $\mathbf{E}$ is the external electric field.
When the atom is in an quasi-resonance radiation field, $\mathbf{E}=\mathbf{E}_0(\mathbf{r}) \cos [\omega t+\varphi(\mathbf{r})]$, where $\omega$ is the frequency of the light, $\mathbf{E}_0$ and $\varphi$ are space-dependent amplitude and phase of the light.
In Rotating Wave Approximation(RWA), the interacting Hamiltonian is
\begin{equation}\label{eq:IntHamRWA}
\begin{split}
  \mathcal{H}_I
     =& -\frac{\hbar}{2}
        \begin{pmatrix}
          \delta                  &    e^{-i\varphi} \Omega \\
          e^{i\varphi} \Omega^*        &   -\delta
        \end{pmatrix}\\
     =& -\frac{\hbar \Omega'}{2}
        \begin{pmatrix}
          \cos \theta             &    e^{-i\varphi} \sin \theta \\
          e^{i\varphi} \sin \theta   &    -\cos \theta
        \end{pmatrix},
\end{split}
\end{equation}
where $\delta=\omega-\omega_{eg}$ is the effective detuning, $\Omega=\mathbf{d}\cdot\mathbf{E}_0$ is the Rabi frequency,  $\Omega'=\sqrt{\delta^2+\Omega^2}$ and $\theta=\arctan (\Omega/\delta)$.
Two eigenstates of $\mathcal{H}_I$ are
\begin{equation}\label{eq:inteignvect}
  \begin{split}
  |\chi_1\rangle =&
    \begin{pmatrix}
      \cos \left(\frac{\theta}{2}\right)\\
      e^{i\varphi} \sin \left(\frac{\theta}{2}\right)
    \end{pmatrix}, \\
  |\chi_2\rangle =&
    \begin{pmatrix}
      -e^{-i\varphi} \sin \left(\frac{\theta}{2}\right)\\
      \cos \left(\frac{\theta}{2}\right)
    \end{pmatrix},
  \end{split}
\end{equation}
with eigenvalues $-\hbar\Omega'/2$ and $\hbar\Omega'/2$ respectively. $(\theta,\varphi)$ is the spherical coordinate of the unit vector on the Bloch sphere corresponding to the spinor.  As shown in Eq.~\ref{eq:IntHamRWA}, the spin is coupled to the effective magnetic field parameterized by, Rabi frequency $\Omega$, phase $\phi$ and effective detuning $\delta$.  By designing proper configuration of the parameters, we may tailor an effective magnetic field.  As the atoms move in the light field adiabatically, their internal states are coherently coupled to the light field, and effective spins are aligned to local magnetic field.

We gave an example of generating steady spin texture of charge $1$ skyrmion by using a Laguerre-Gaussian beam.  Assume that the beam is propagating in $z$-direction
\begin{equation}\label{eq:LaguerreGaussianBeam}
  \begin{split}
    \mathbf{E}=&\Re\Bigg\{\mathbf{e} e^{i(kz-\omega t)} \frac{D}{\sqrt{1+z^2/z_R^2}}
    {\left[\frac{r\sqrt{2}}{w(z)}\right]}^l L_p^l\left(\frac{2r^2}{w^2(z)}\right)\\
    &\times\exp\left[-\frac{r^2}{w^2(z)}\right]
    \exp\left[-\frac{ikr^2z}{2(z^2+z_R^2)}\right]
    \exp\left(-il\phi\right)\\
    &\times\exp\left[i(2p+l+1)\arctan \left(\frac{z}{z_R}\right)\right]\Bigg\},
  \end{split}
\end{equation}
where $(r,\phi,z)$ is the cylindric coordinate, $\mathbf{e}$ is the polarization vector in $x$-$y$ plane, $w(z)$ is the radius of the beam, $z_R$ is the Rayleigh range, $L_p^l(x)$ is the associated Laguerre polynomial, $\omega$ is the frequency of the beam red-detuned to the resonance, $k$ is the wave vector and $D$ is a constant.
\begin{figure}
  \centering
  \includegraphics[width=0.43\textwidth]{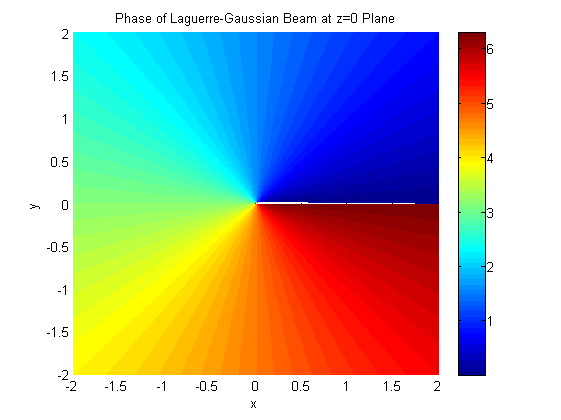}\\
  \includegraphics[width=0.43\textwidth]{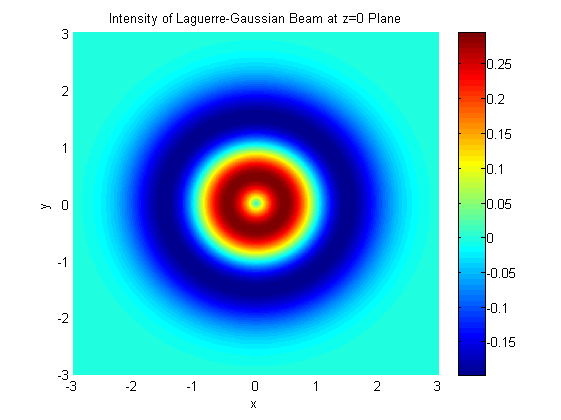}\\
  \caption{Phase and Intensity of Laguerre-Gaussian Beam at $z=0$ Plane}\label{fig:LGbeam}
\end{figure}

\begin{figure}
  \centering
  \includegraphics[width=0.48\textwidth]{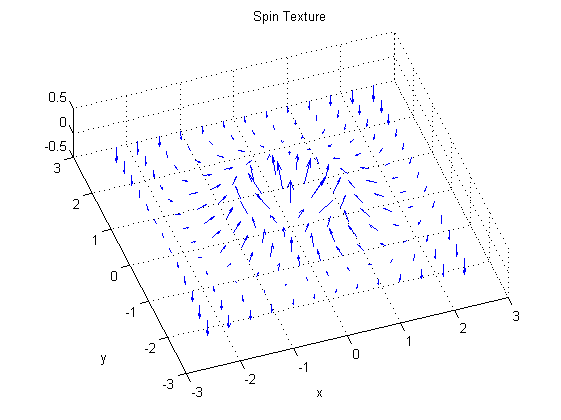}\\
  \caption{Spin Texture Induced by Laguerre-Gaussian Beam}\label{fig:spintexture}
\end{figure}

The atom is confined in 2D plane at $z=0$.  By definition
the Rabi frequency is
\begin{equation}\label{eq:RabiFreqLaguerre}
  \begin{split}
    e^{-i\varphi}\Omega=&\tilde{D}r\left(1-\frac{r^2}{w^2(0)}\right) \exp\left[-\frac{r^2}{w^2(0)}\right]
    \exp\left(-i\phi\right),
  \end{split}
\end{equation}
where $\tilde{D}$ is a constant.
At the origin, $\Omega$ vanishes, red-detuned frequency orientate the spinor downwards.  Along the radial direction, $\Omega$ increases tilting the orientation to $x$-$y$ plane, and phase $\phi$ orient the tilting along the radial direction.  At $\infty$, $\Omega$ vanishes again, red-detuned frequency redirect the spinor downwards, which is topologically trivial.
To flip the spin, we need ``shifting'' the laser to blue detuning.  It is hard to realize steady light field with space dependent frequency, however effective detuning $\delta$ can also be shifted by Zeeman effect.  Therefore if we choose the internal state with non-vanishing magnetic moment, we may apply a real magnetic field to shift down the energy level $|e\rangle$ instead of shifting laser frequency.   The consequent spin texture is shown in Fig.~\ref{fig:spintexture}, which is a realization of skyrmion with charge $c=-1$.


Assuming that initially, atom is in state $|\chi_1\rangle$, the Berry connection and curvature of this submanifold is
\begin{equation}\label{eq:BeryyCurvature}
\begin{split}
{{\mathbf{A}}}=& i \langle\chi_1 |\nabla |\chi_1\rangle = - \sin^2\left(\frac{\theta}{2}\right) \nabla \phi,\\
\mathbf{B}=& \nabla \times \mathbf{A} = - \frac{1}{2} \sin\theta \nabla\theta \times \nabla\phi.
\end{split}
\end{equation}
Integrating over the whole plane, we can calculate the topological
charge
\begin{equation}\label{eq:TopoCharge}
\begin{split}
C=\frac{1}{2\pi}\int
\mathbf{B} \cdot d\mathbf{S} = - \frac{1}{4\pi}  \int \sin\theta d\theta \wedge d\phi = -1.
\end{split}
\end{equation}
This spin texture also induces another effective magnetic field due to Berry curvature. if the atom turn around the origin along the circle where the laser is on resonance, it will picks up a phase factor of $e^{\pm i\pi}=-1$.

In adiabatic limit, we may use a beam arbitrary weak.  If the motion of the atom can be considered as adiabatic, the gap between ground and excited submanifold should be much larger than kinetic energy scale, which set the lower limit of the beam Intensity.  On the other hand, strong beam would cause heating due to spontaneous emission. To reduce the spontaneous emission, we may choose long life time internal state, use two-photon Raman transition coupling to internal state in the ground level, or use the dark state.  


In conclusion, we proposed a scheme to generate steady topologically non-trivial artificial spin texture in cold atom systems.  We proposed a scheme of Laguerre-Gaussian beam to induce spin texture corresponding to a skyrmion of charge $c=\pm1$.  It is also possible to generate more complex spin texture by choosing proper laser mode in this scheme. Combining with the techniques of artificial magnetic field, it provides a steady method to imprint spin texture to quantum Hall states in cold atoms.  It will be interesting to investigate particle's motions in a topologically non-trivial spin texture.

The author would thank Benoit Dou\c{c}ot for his helpful discussions at Les Houches summer school.  The early studies Ref.~\citenum{Zhai2003} and~\citenum{Leslie2009} were pointed out by Jean Dalibard to author.

\end{document}